**Letter**

# Bayesian Unmixing using Sparse Dirichlet Prior with Polynomial Post-nonlinear Mixing Model

Fahime Amiri * Non-member of IEEJ, Mohammad Hossein Kahaei * a) Non-member of IEEJ

A sparse Dirichlet prior is proposed for estimating the abundance vector of hyperspectral images with a nonlinear mixing model. This sparse prior is led to an unmixing procedure in a semi-supervised scenario in which exact materials are unknown. The nonlinear model is a polynomial post-nonlinear mixing model that represents each hyperspectral pixel as a nonlinear function of pure spectral signatures corrupted by additive white noise. Simulation results show more than 50% improvement in the estimation error.

## 1. Introduction

Hyperspectral unmixing is a post-processing technique that recognizes pure spectral signatures and their corresponding proportions in Hyperspectral Images (HSI), referred to as endmembers and abundances, respectively [1]. Under linear mixture assumption, a hyperspectral pixel is expressed as a convex mixture of some endmembers resulting in the Linear Mixing Model (LMM). Due to encourage the drawbacks of LMMs for the images of the materials like sands or trees [1], the nonlinear mixing models are presented. The Polynomial Post-Nonlinear Mixing Model (PPNMM) is one of the most interesting nonlinear mixing models used for supervised hyperspectral unmixing [2]. Although, an unsupervised version of PPNMM unmixing has been presented in [3], a third-party Endmember Extraction Algorithm (EEA) is necessary yet.

In this paper, we propose the Sparse Dirichlet Prior with PPNMM (SDP-PPNMM) algorithm in a semi-supervised manner that means we do not need any EEA and the lack of knowledge about pure endmembers is compensated just by selecting suitable priors. In fact, we extend our previous work [4] to a nonlinear mixing model. We assume that a large library of endmembers is available, which is a realistic assumption due to collecting a wide variety of spectral signature of various common materials during few decades [5].

It is important to note that due to contribution of only a small number of endmembers of an extremely large library in each hyperspectral pixel, the abundance vector could be sparse. Accordingly, in this paper we consider this case for estimating abundance vectors in a Bayesian sense. In this way, unmixing and endmember selection from a large library are performed simultaneously. The Markov Chain Monte Carlo (MCMC) sampler used to generate samples based on derived posterior.

To elaborate, in a nonlinear mixing model, a hyperspectral pixel is defined as a nonlinear function of a linear mixture of endmember signatures affected by noise term as

$$\mathbf{y} = \boldsymbol{g}\left(\textstyle\sum_{r=1}^{R} a_r \mathbf{m}_r\right) + \mathbf{n} = \boldsymbol{g}(\mathbf{Ma}) + \mathbf{n}, \quad (1)$$

a) Correspondence to: Mohammad Hossein Kahaei. E-mail: kahaei@iust.ac.ir

* School of Electrical Engineering, Iran University of Science and Technology, Tehran, Iran.

where $\mathbf{y}$ is an L-dimensional hyperspectral pixel, $\mathbf{m}_r$ is the spectral signature of the $r$th endmember in the library $\mathbf{M}$, $a_r$ is the corresponding abundance, $R$ is the number of endmembers in the library, $\boldsymbol{g}$ is a nonlinear transformation, and $\mathbf{n}$ is additive white Gaussian noise with zero-mean and variance $\sigma^2$.

Here, a second-order polynomial is employed for the nonlinear function. It has been shown that second order polynomial is an appropriate approximation for nonlinear mixing models [6], since higher order terms are negligible and could be merged in noise term. A PPNMM [2], is given by

$$\boldsymbol{g}_b:\ (\mathbf{a}, b) \rightarrow \mathbf{Ma} + b(\mathbf{Ma}) \odot (\mathbf{Ma}) \quad (2)$$

where $\odot$ is the Hadamard product. This model contains both bilinear and linear models. According to [2], using $b$ as a single amplitude parameter for the nonlinear term, lower complexity is achieved.

## 2. Bayesian Framework

We utilize the hierarchical Bayesian solution [2]. Our motivation is to select a proper prior for the abundance vector which leads to not using any EEA. Then, the likelihood function and joint conditional pdf's are computed based on the proposed priors.

The likelihood function of the mixed pixel is denoted as

$$f(\mathbf{y}|\mathbf{a}, b, \sigma^2) = \left(\frac{1}{2\pi\sigma^2}\right)^{\frac{L}{2}} \exp\left(-\frac{\|\mathbf{y}-\boldsymbol{g}_b(\mathbf{Ma})\|^2}{2\sigma^2}\right) \quad (3)$$

in which $\mathbf{a}, b, \sigma^2$ should be estimated. Due to the two physical constraints, sum-to-one and non-negativity of the abundance vector, we propose the sparse symmetric Dirichlet distribution [7] as a prior for the abundance vector a as

$$f(\mathbf{a}) \sim \mathcal{D}(\mathbf{a}; \beta) = \frac{\Gamma(\beta R)}{\Gamma(\beta)^R} \prod_{r=1}^{R} a_r^{\beta-1} \quad (4)$$

where $\Gamma(.)$ is the Gamma function and $\beta$ exhibits the concentration parameter. This distribution presents a sparse behavior for $\beta < 1$ and corresponds to a uniform distribution over the standard $(R-1)$-simplex for $\beta = 1$. The latter case is commonly used in unmixing problems [2]. Note that this case is limited to supervised unmixing applications in which the exact endmembers must be known. In more realistic scenarios, however, exact mixing endmembers are unknown and only a large spectral library is given and the concentration parameter plays an essential role. Here, accordingly we consider $\beta < 1$ and show that this case is suitable for a wide range of applications. For the unknown parameter $\sigma^2$ the Jeffrey's prior is assigned. A normal distribution with zero-mean and variance $\sigma_b^2$ is assigned to the nonlinear parameter $b$, where $\sigma_b^2$ is a hyperparameter for which the Inverse-Gamma prior with parameters $(\gamma, \nu) = (1, 0.01)$ according to [2].

Based on the Bayes theorem, the joint distribution of the unknown parameters $\boldsymbol{\theta} = \{\mathbf{a}, b, \sigma^2, \sigma_b^2\}$ is described as

$$f(\boldsymbol{\theta}|\mathbf{y}) \propto f(\mathbf{y}|\boldsymbol{\theta}) f(\mathbf{a}, b, \sigma^2|\sigma_b^2) f(\sigma_b^2) \cdots\cdots (5)$$

The derived distribution is too complex to utilize the MMSE or MAP estimators. Thus, we use the Metropolis-Within-Gibbs sampler to sample data according to the conditional distributions. The conditional pdf of parameters are computed as:

$$f(a_r|\mathbf{y}, \boldsymbol{\theta}_{\backslash a_r}) \propto \exp\left(-\frac{\|\mathbf{y}-\boldsymbol{g}_b(\mathbf{Ma})\|^2}{2\sigma^2}\right) \prod_{r=1}^{R} a_r^{\beta-1} \cdots\cdots (6)$$

$$b|\mathbf{y}, \boldsymbol{\theta}_{\backslash b} \sim \mathcal{N}\left(\frac{\sigma_b^2(\mathbf{y}-\mathbf{Ma})^T\boldsymbol{h}(\mathbf{a})}{\sigma_b^2\boldsymbol{h}(\mathbf{a})^T\boldsymbol{h}(\mathbf{a})+\sigma^2}, \frac{\sigma_b^2\sigma^2}{\sigma_b^2\boldsymbol{h}(\mathbf{a})^T\boldsymbol{h}(\mathbf{a})+\sigma^2}\right) \cdots\cdots (7)$$

$$\sigma^2|\mathbf{y}, \boldsymbol{\theta}_{\backslash \sigma^2} \sim \mathcal{IG}\left(\frac{L}{2}, \frac{\|\mathbf{y}-\boldsymbol{g}_b(\mathbf{Ma})\|^2}{2}\right) \cdots\cdots (8)$$

$$\sigma_b^2|\mathbf{y}, \boldsymbol{\theta}_{\backslash \sigma_b^2} \sim \mathcal{IG}\left(\frac{1}{2}+\gamma, \frac{b^2}{2}+\nu\right) \cdots\cdots (9)$$

After initialization, in each iteration first sample $a_r$ by MCMC algorithm, then the coefficient $b$, after that $\sigma^2$ and finally $\sigma_b^2$.

## 3. Experimental Result

We evaluate the SDP-PPNMM algorithm on a synthetic data. The results are compared to that of the classical uniform prior. A synthetic hyperspectral pixel composed of two endmembers is generated. We select 6 endmembers randomly from the USGS library [6] and make our own library. According to (2), a synthetic pixel is generated by mixing the endmembers using $a = [0.3, 0.7, 0, 0, 0, 0]$ and $b = 0.2$. As seen, the abundance vector contains 4 zero elements which let us simulate a semi-supervised scenario. In this case we are not aware of neither the number of endmembers nor the associated ones in the mixing process. We also choose $\beta = 0.5$ making a sparse distribution. Simulations are run 20 times with 10000 MCMC and 1000 burn-in iterations.

The posterior distribution of the 1st, 2nd, and 3rd abundance values are illustrated in Figs. 1-a, b, and c, respectively. As seen in Fig. 1a, the first abundance value estimated by the proposed SDP-PPNMM method is $\hat{a}_1 = 0.3011$ which is very close to the real value 0.3. This value for [2] is equal to 0.2806. In Fig. 1b one can see that although both algorithms achieve almost a similar accuracy for $a_2$, the SDP-PPNMM algorithm clearly outperforms the PPNMM for $a_3$. Note that the real values of the 3rd to 6th elements of the abundance vector are zero for which we have only shown the estimated posterior for the 3rd one. It is seen that the estimated pdf by the SDP-PPNMM is sharper and more concentrated on zero than that of [2] which leads to more accurate estimates. To evaluate the estimation error (MSE) and the reconstruction error (RE) quantitatively, we define the following parameters where listed in Table 1.

$$\mathrm{MSE} = \frac{1}{P}\sum_{p=1}^{P}\left\|\hat{\mathbf{a}}_p - \mathbf{a}_p\right\|^2, \ \mathrm{RE} = \sqrt{\frac{1}{PL}\sum_{p=1}^{P}\left\|\hat{\mathbf{y}}_p - \mathbf{y}_p\right\|^2} \cdots (10)$$

Table 1. Estimation and reconstruction errors $\times 10^{-2}$

| Algorithm | MSE | RE |
|---|---|---|
| SDP-PPNMM | **0.0238** | **5.17** |
| PPNMM [2] | 0.0543 | 5.30 |

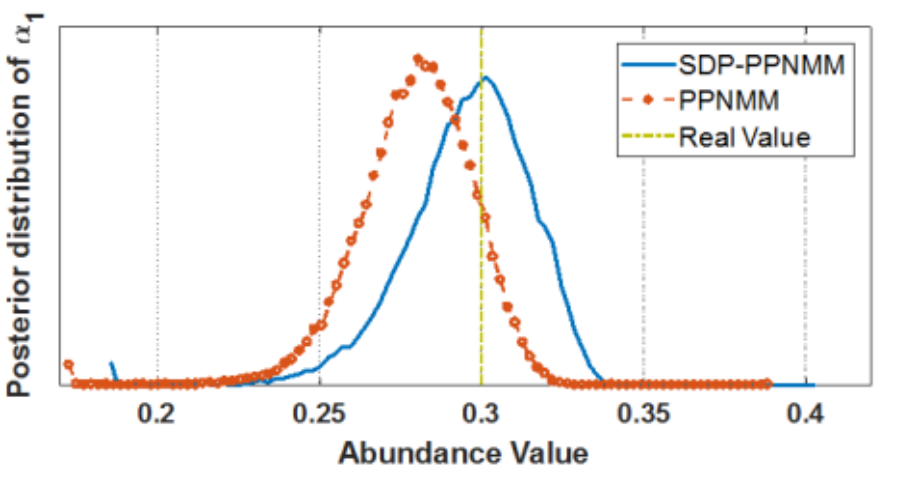

(a) First estimated abundance value

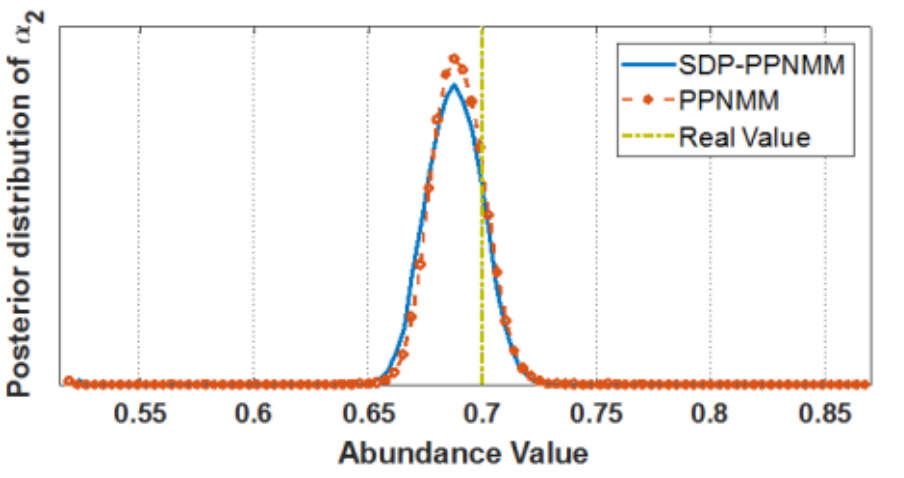

(b) Second estimated abundance value

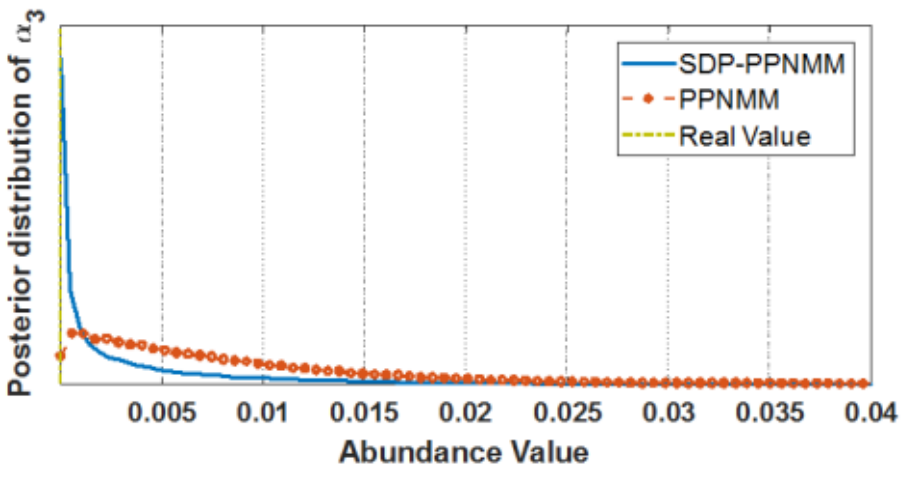

(c) Third estimated abundance value

Fig. 1. Estimated posterior distributions of the abundance vector for the SDP-PPNMM and [2]

## 4. Conclusion

We derived a hierarchical Bayesian algorithm for unmixing of hyperspectral images based on the PPNMM. A Dirichlet prior was proposed for modeling the abundance vector sparsity. We set the concentration parameter in such way that the abundance pdf leads to a sparse distribution. In this way, if a huge library is given, the unmixing procedure could be done precisely and any third-party EE algorithm would not be necessary. The MCMC method as used to estimate the posterior. By applying the sparse Dirichlet prior to the mixed pixel, the absolute error of the estimated abundance vector became smaller than half of that of the uniform prior.

## References

(1) Keshava N., Mustard J. F., "Spectral unmixing," IEEE Signal Processing Magazine, 19, no. 1, pp. 44-57 (2002)

2. Altmann Y., Halimi A., Dobigeon N., and Tourneret J.: 'Supervised nonlinear spectral unmixing using a postnonlinear mixing model for hyperspectral imagery',IEEE Trans. Image Process., 21, (6), pp. 3017–3025 (2012)

3. Altmann Y., Dobigeon N., and Tourneret J-V.: 'Unsupervised Post-Nonlinear Unmixing of Hyperspectral Images Using a Hamiltonian Monte Carlo Algorithm', IEEE Trans. Image Process. 23, (6), pp: 2663-2675 (2014)

4. Amiri F., Kahaei M. H., "New Bayesian approach for semi-supervised hyperspectral unmixing in linear mixing models", 25th Iranian Conference on Electrical Engineering (ICEE), pp. 1752-1756 (2017)

5. Clark R., Swayze G., Wise R., Livo E., Hoefen T., Kokaly R., and Sutley S. 'USGS digital spectral library splib06a: U.S. Geological Survey, Digital Data Series 231' http://speclab.cr.usgs.gov/spectral.lib06, (2007)

6. Nascimento J. M. P. and Bioucas-Dias J. M., SPIE, "Nonlinear mixture model for hyperspectral unmixing," in Proc. SPIE Image Signal Process. Remote Sens., 7477, (1), p. 747 70I (2009)

7. Ng K. W., Tian G. L., Tang and M. L., 'Dirichlet and Related Distributions: Theory, Methods and Applications', (Wiley Press, New York, 2011)